\documentclass[%
 reprint,
 amsmath,amssymb,
 aps,
 prb,
]{revtex4-1}

\usepackage{graphicx,overpic}
\usepackage{dcolumn}
\usepackage{bm}
\usepackage{hyperref}
\usepackage{amsmath}
\usepackage{color}
\usepackage{comment}

\begin{document}


\title{Geodesics in the (anti-)de Sitter spacetimes}

\author{Nguyen Phuc Ky Tho}
\email{tho.nguyenphucky@hust.edu.vn}
\affiliation{
Department of Theoretical Physics, Hanoi University of Science and Technology (HUST)\\
No. 1 Dai Co Viet Road, Hanoi, 100000, Vietnam
}

\date{\today}%

\begin{abstract}
A class of exact solutions of the geodesic equations in (anti-)de Sitter AdS$_4$ and dS$_4$ spacetimes is presented. The geodesics for test particles in AdS$_4$ and dS$_4$ spacetimes are respectively sinusoidal and hyperbolic sine world lines. The world line for light rays is straight lines as known. In particular, the world lines of test particles are not dependent on their energy. Spontaneous symmetry breaking of AdS$_4$ spacetime provides a physical explanation for arising of the virtual particle and antiparticle pairs in vacuum. Interestingly, the energy of a pair and the time its particles moving along their geodesics can be related by a relation similar to Heisenberg uncertainty one pertaining quantum vacuum fluctuations. The sinusoidal geodesics of AdS$_4$  spacetime can describe the world lines of the virtual particles and antiparticles. The hyperbolic sine geodesics of dS$_4$ spacetime can explain why galaxies move apart with positive accelerations.
\end{abstract}

\maketitle


\section{Introduction}

The Einstein equations can be written in the following form
\begin{equation} \label{field_eqn}
R_{\mu\nu} - \frac{1}{2} Rg_{\mu\nu} + \lambda g_{\mu\nu} = 8\pi G T_{\mu\nu}^M \quad (g_{00} = 1, c = 1)
\end{equation}
where $R_{\mu\nu}$ is the Ricci tensor, $T_{\mu\nu}^M$ is the stress-energy tensor. The cosmological constant $\lambda$ is the free parameter. \cite{weinberg1989,carroll1992,carroll2001,padmanabhan2003,peebles2003} Neglecting the stress-energy tensor of the matter, we obtain an empty space with the vacuum energy density $\rho_{vac}=\lambda/8\pi G$. Hence the Einstein equations will become
\begin{equation} \label{f0_eqn}
R_{\mu\nu} - \frac{1}{2} Rg_{\mu\nu} + \lambda g_{\mu\nu} = 0.
\end{equation}
The indices of the above equations are contracted, giving the Ricci scalar $R=4 \lambda$.
The equations (\ref{f0_eqn}) will then get the form
\begin{equation} \label{f0c_eqn}
R_{\mu\nu} = \lambda g_{\mu\nu}.
\end{equation}
The most important vacuum solution of the equations (\ref{f0c_eqn}) is the Schwarzschild--(anti-)de Sitter spacetime \cite{bengtsson1998,hackmann2015,hackmann2008,hackmann2010,cruz2005}
\begin{equation} \label{sss_eqn}
\begin{split}
ds^2 &= (1 - 2GM/r - \lambda r^2/3) dt^2 \\
     &\quad - \frac{dr^2}{(1 - 2GM/r - \lambda r^2/3)} - r^2 d\Omega^2
\end{split}
\end{equation}
where $d\Omega^2 = d\theta^2 + \sin^2 \theta d\phi^2$ is the metric of a two-sphere.
For $\lambda = 0$, we obtain the Schwarzschild spacetime where the curvature is caused by the spherical mass $M$ at the origin of the coordinate system.
For $M = 0$ and $\lambda > 0$, putting $\lambda = 3\omega^2 > 0$, the metric (\ref{sss_eqn}) reduces to the de Sitter spacetime (dS$_4$) with a false singularity $r_s = 1/\omega$, and dS$_4$ is an ``empty space'' with $\rho_{vac}=3\omega^2/8\pi G > 0$
\begin{equation} \label{ss1_eqn}
\begin{split}
ds^2 &= (1 - \omega^2 r^2) dt^2 \\
     &\quad - \frac{dr^2}{(1 - \omega^2 r^2)} - r^2 d\Omega^2.
\end{split}
\end{equation}
For $M = 0$ and $\lambda < 0$, putting $\lambda = -3\omega^2 < 0$, we have the metric of the anti-de Sitter (AdS$_4$) spacetime that has not any singularity, and AdS$_4$ is an ``empty space'' with $\rho_{vac}=-3\omega^2/8\pi G < 0$
\begin{equation} \label{ss0_eqn}
\begin{split}
ds^2 &= (1 + \omega^2 r^2) dt^2 \\
     &\quad - \frac{dr^2}{(1 + \omega^2 r^2)} - r^2 d\Omega^2.
\end{split}
\end{equation}
The change from the de Sitter spacetime to the anti-de Sitter spacetime is performed by replacing $\omega \rightarrow i\omega$ and vice versa. These spacetimes differ from the Minkowski spacetime, which is really an empty space with $\rho_{vac}=0$.
Note that in cosmology, the positive and negative cosmological constant $\lambda$ are respectively considered candidates for dark energy and dark matter. \cite{burdyuzha2013,ohanian1994}

In this study, we present a class of exact solutions of the geodesic equations in the AdS$_4$ and dS$_4$ spacetimes. When considering the radial motion of the test particles, the equations are significantly simplified, as far as the simple harmonic oscillator one, and the solutions become readily obtained, in surprisingly fundamental forms of sinusoidal and hyperbolic sine functions. The mathematical derivation is given in Sec.~\ref{sec:geo}. Our attempt to explain physics of the obtained results is given in Sec.~\ref{sec:phys}, where again fundamental problems of the vacuum energy and cosmological constant are being put forth.

\section{The geodesics in the (anti-)de Sitter spacetimes} \label{sec:geo}

\begin{figure}[!t]
\centering
\includegraphics[width=0.22\textwidth, angle=0]{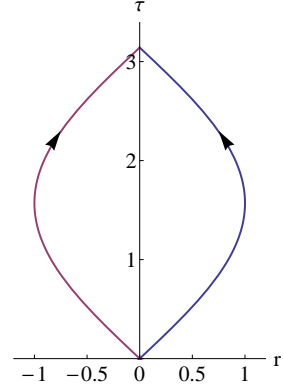}
\caption{AdS$_4$ sinusoidal worldline for $\omega = 3 \times 10^{21} s^{-1}$, i.e. $\lambda \sim 3 \times 10^{43} s^{-2}$, typical values for a virtual electron-positron pair. The units of $\tau$ and r are $10^{-21} s$ and $3 \times 10^{-11} cm$, respectively.}
\label{fig1}
\end{figure}

First the metric tensor of AdS$_4$ spacetime, from (\ref{ss0_eqn}), is put into the following form
\begin{equation} \label{gab_eqn}
g_{\alpha \beta} = 
\begin{pmatrix}
        e^A & 0 & 0 & 0 \\
        0 & -e^B & 0 & 0 \\
        0 & 0 & -r^2 & 0 \\
        0 & 0 & 0 & -r^2 sin^2 \theta
\end{pmatrix}
\end{equation}
where $e^{A(r)} = 1 + \omega^2r^2; e^{B(r)} = 1/(1 + \omega^2r^2)$. 
The Christoffel symbols for the metric tensor have been given
\begin{equation} \label{chris_eqn}
\begin{array}{lll}
\Gamma_{01}^0 = \Gamma_{10}^0 = \frac{1}{2}A', &\Gamma_{00}^1 = \frac{1}{2}A' e^{A-B}, 
&\Gamma_{11}^1 = \frac{1}{2}B' \\
\Gamma_{22}^1 = -r e^{-B}, &\Gamma_{33}^1 = -r \sin^2 \theta e^{-B} \\
\Gamma_{12}^2 = \Gamma_{21}^2 = 1/r, &\Gamma_{33}^2 = -\sin \theta \cos \theta \\
\Gamma_{13}^3 = \Gamma_{31}^3 = 1/r, &\Gamma_{23}^3 = \Gamma_{32}^3 = \cot \theta
\end{array}
\end{equation}
where $A'=\frac{dA}{dr}=\frac{2 \omega^2 r}{1 + \omega^2 r^2}; B'=\frac{dB}{dr}=-\frac{2 \omega^2 r}{1 + \omega^2 r^2}$. \\
The geodesic equations are given
\begin{equation} \label{geo_eqn}
\frac{d^2 x^{\mu}}{d \tau^2}+\Gamma_{\alpha \beta}^{\mu} \frac{dx^{\alpha}}{d \tau} \frac{dx^{\beta}}{d \tau} = 0
\end{equation}
and the following quantity is a constant of the motion
\begin{equation} \label{gab01_eqn}
g_{\alpha \beta}\frac{dx^{\alpha}}{d \tau} \frac{dx^{\beta}}{d \tau} = 
\begin{cases}
1 \quad \text{for massive particles} \\
0 \quad \text{for light}
\end{cases}
\end{equation}
The geodesic equations (\ref{geo_eqn}) can be written as follows
\begin{eqnarray} \label{geos1_eqn}
\ddot{t} + A' \dot{r} \dot{t} = 0 \\
\ddot{r} + \frac{1}{2}A' e^{A-B}(\dot{t})^2 + \frac{1}{2}B'(\dot{r})^2 \nonumber \\
- r e^{-B}(\dot{\theta})^2 - r \sin^2 \theta e^{-B}(\dot{\phi})^2 = 0 \\
\ddot{\theta} + \frac{2}{r} \dot{r} \dot{\theta} - \cos \theta \sin \theta (\dot{\phi})^2 = 0 \\
\ddot{\phi} + \frac{2}{r} \dot{r} \dot{\phi} + 2 \cot \theta \dot{\phi}\dot{\theta} = 0
\end{eqnarray}
where $\dot{r}=\frac{dr}{d\tau}, \dot{t}=\frac{dt}{d\tau}, \ddot{r}=\frac{d^2 r}{d \tau^2}, \ddot{t}=\frac{d^2 t}{d \tau^2},$ and so on. We can restrict to the equatorial plane $\theta=\pi /2$  because of angular momentum conservation. Then we have only three following geodesic equations
\begin{eqnarray} \label{geos2_eqn}
\ddot{t} + \frac{2 \omega^2 r}{1 + \omega^2 r^2} \dot{t}\dot{r} = 0 \label{geos2t_eqn} \\
\ddot{r} + \omega^2 r (1 + \omega^2 r^2) \dot{t}^2 - \frac{\omega^2 r}{1 + \omega^2 r^2} \dot{r}^2 \nonumber \\
- r (1 + \omega^2 r^2) \dot{\phi}^2 = 0 \label{geos2r_eqn} \\
\ddot{\phi} + \frac{2}{r} \dot{r} \dot{\phi} = 0 \label{geos2p_eqn}
\end{eqnarray}
The relation (\ref{gab01_eqn}) becomes
\begin{equation} \label{gab012_eqn}
(1 + \omega^2 r^2)\dot{t}^2 - \frac{\dot{r}^2}{1 + \omega^2 r^2} = 
\begin{cases}
1 \quad \text{for massive particles} \\
0 \quad \text{for light}
\end{cases}
\end{equation}
By integrating equations (\ref{geos2t_eqn}) and (\ref{geos2p_eqn}) we have
\begin{equation} \label{tdot_eqn}
\dot{t} = \frac{E}{1 + \omega^2 r^2}
\end{equation}
and
\begin{equation}
\dot{\phi} = \frac{L}{r^2}
\end{equation}
where $E=const$ and $L=const$, the energy and angular momentum of the test particle. 
From the result (\ref{gab012_eqn}) for massive particles and the equation (\ref{tdot_eqn}), we can find the following relation
\begin{equation} \label{rdot2_eqn}
\dot{r}^2 = E^2 - (1 + \omega^2 r^2)
\end{equation}
Without loss of generality, by supposing $\phi = const\ (0 \leq \phi \leq 2\pi)$ we restrict to the radial motion of the test particle, and its angular momentum $L=r^2 \dot{\phi} = 0$. 
By inserting the equations (\ref{tdot_eqn}) and (\ref{rdot2_eqn}) into the equation (\ref{geos2r_eqn}) we obtain the equation for a simple harmonic oscillator
\begin{equation} \label{hor_eqn}
\ddot{r} + \omega^2 r = 0
\end{equation}
and the solutions
\begin{equation} \label{rads_eqn}
r = \pm \frac{1}{\omega} \sin(\omega \tau)
\end{equation}
with null initial conditions.
The world lines in the AdS$_4$ spacetime are therefore sinusoidal (see Fig.~\ref{fig1}).
Next we obtain $\dot{r}^2 = E^2$ for light rays from the result (\ref{gab012_eqn}) and the equation (\ref{tdot_eqn}). Then from the equations (\ref{geos2r_eqn}) and (\ref{gab012_eqn}) we find the equation of motion $\ddot{r} = 0$ with solutions
\begin{equation} \label{cone_eqn}
r = \pm \tau.
\end{equation}
Thus the world line for light rays is straight lines as known.
Finally, the geodesic equation in the dS$_4$ metric $(\lambda > 0)$ will be given from the equation (\ref{hor_eqn}) by the substitution $\omega \rightarrow i \omega$
\begin{equation} \label{shr_eqn}
\ddot{r} - \omega^2 r = 0
\end{equation}
with the solutions
\begin{equation} \label{rds_eqn}
r = \pm \frac{1}{\omega} \sinh (\omega \tau).
\end{equation}
The world lines in the dS$_4$ spacetime are hyperbolic sine lines.
An interesting result is that the world line functional forms (\ref{rads_eqn}), (\ref{cone_eqn}) and (\ref{rds_eqn}) show no dependence on the energy $E$ of the test particle.

We see that the metrics (\ref{ss1_eqn}) and (\ref{ss0_eqn}) do not depend explicitly on the time-coordinate $t$. These metrics are called static. Now we derive the time-dependent dS$_4$ metric with the flat slicing coordinates from the static metric (\ref{ss1_eqn}).\cite{ripken2013} Considering the following coordinate transformations
\begin{equation}
r=r' e^{\omega t'}, \quad t=t' - \frac{1}{2\omega} ln(r'^2 e^{2\omega t'} - \frac{1}{\omega^2})
\end{equation}
where $d\Omega^2$ is invariant. The simple calculation demonstrates that
\begin{equation}
(1-\omega^2 r^2)dt^2 - \frac{dr^2}{1-\omega^2 r^2} = dt'^2 - e^{2\omega t'}dr'^2.
\end{equation}
Hence the static dS$_4$ metric reduces to the time-dependent dS$_4$ metric with the flat slicing coordinates
\begin{equation} \label{dsmp_eqn}
ds^2=dt'^2 - e^{2\omega t'}(dr'^2 + r'^2 d\Omega^2).
\end{equation}
After discarding the primes, we get the time-dependent de Sitter metric
\begin{equation} \label{dsm_eqn}
ds^2=dt^2 - e^{2\omega t}(dr^2 + r^2 d\theta^2 + r^2 \sin^2 \theta d\phi^2).
\end{equation}
The time-dependent AdS$_4$ metric can be given by the substitution $\omega \rightarrow i\omega$
\begin{equation} \label{adsm_eqn}
ds^2=dt^2 - e^{2i\omega t}(dr^2 + r^2 d\theta^2 + r^2 \sin^2 \theta d\phi^2).
\end{equation}
This complex metric does not evidently describe a real spacetime, hence its symmetry certainly must be broken in the real world when $e^{2i\omega t}=+1$. Then we get a relation $\omega t=n\pi$ where $n=0,\pm 1, \pm 2,$ and so on. Thus the anti-de Sitter spacetime (\ref{adsm_eqn}) reduces to the Minkowski spacetime at the following moments
\begin{equation}
t_n=\frac{n\pi}{\omega}
\end{equation}
and we get the relation
\begin{equation}
\omega \Delta t = \pi
\end{equation}
where $\Delta t=t_{n+1}-t_n=\pi/\omega$. Multiplying two sides of the above relation by the Planck constant, we obtain a relation similar to Heisenberg uncertainty one pertaining quantum vacuum fluctuations
\begin{equation}
\varepsilon \Delta t=\pi \hbar \approx \hbar
\end{equation}
where $\varepsilon=\hbar \omega$. Therefore spontaneous symmetry breaking of AdS$_4$ spacetime provides a physical explanation for arising of some particle with an energy $\varepsilon$ in the vacuum. However, which particle does have such energy $\varepsilon$, and how does it moves?

\section{The physical explanations} \label{sec:phys}

As we know that the quantum vacuum is a very active place, filled with virtual particle and antiparticle appearing and disappearing continuously. These virtual particles and antiparticles appear and disappear everywhere and in all directions. We may ascribe $\varepsilon=\hbar \omega=\frac{\hbar \omega}{2}+\frac{\hbar \omega}{2}$ to the total energy of the virtual particle and antiparticle pair. The virtual particle which emerges from the vacuum moves along the geodesic in a time interval $\Delta t \sim 1/\omega$. The geodesics of virtual particle and antiparticle are the sinusoidal world lines (\ref{rads_eqn}) in the AdS$_4$ spacetime. Every virtual particle in the AdS$_4$ vacuum has an energy $\frac{\hbar \omega}{2}$, where $\omega$ can take any value $(0\leq \omega <+\infty)$, then the vacuum energy density in the AdS$_4$ spacetime can be defined by\cite{moffat2009}
\begin{equation}
\rho_{vac}=\frac{1}{V} \sum_k \frac{\hbar \omega_k}{2} \approx \frac{\hbar}{2\pi^2}\int_0^{\omega_{max}} \omega^3 d\omega \propto \hbar \omega_{max}^4.
\end{equation}
This vacuum energy density is strongly divergent with an ultraviolet cut-off frequency $\omega_{max}$. 
Thus the vacuum energy density diverges not only in quantum field theory but also in general relativity. However the negative vacuum energy density $\rho_{vac}=-3\omega^2/8\pi G < 0$ can not cause any gravitational effect from the viewpoint of general relativity. If the AdS$_4$ spacetime can describe a vacuum with great energy density values in the smallest scales, then the dS$_4$ spacetime can represent a vacuum with small energy density values in the greatest scales. A vacuum energy density of the present universe,\cite{marsh2008,oldershaw2009} $\rho_{vac}\propto 10^{-29} g/cm^3$, is compatible totally with a cosmological constant $\lambda \sim 10^{-35} s^{-2}$, however it does not have any relation with the quantum vacuum energy density. Thus is it not necessary to cancel the famous discrepancy of 120 orders of magnitude between the theoretical and observational values of the cosmological constant?

The sinusoidal geodesics in AdS$_4$ spacetime can describe the world lines of the virtual particles and antiparticles which appear and disappear continuously in the vacuum. 
The hyperbolic sine geodesics (\ref{rds_eqn}) in dS$_4$ spacetime with $\lambda > 0$ can explain why galaxies move apart with positive accelerations. Indeed, from the equation (\ref{shr_eqn}) we get a positive acceleration of the test particle $\ddot{r}/r=\omega^2 > 0$. It is easy to find the following relation
\begin{equation}
\frac{\nu_2}{\nu_1} =  \frac{\tau_1}{\tau_2} = \frac{arcsinh(\omega r_1)}{arcsinh(\omega r_2)} \approx \frac{r_1}{r_2} < 1 \rightarrow \nu_2 < \nu_1
\end{equation}
that implies the red shift.
Note that values of the positive cosmological constant might be very large in the Planck era or in the vacuum of high energy physics, but its present observational value is very small. Such large values of the cosmological constant may be necessary to explain a fast expansion of the universe (an inflation). Does the cosmological constant decay in the evolution of the universe?

\section{Summary}

We have considered the geometry properties of the (anti-)de Sitter spacetimes, determined by the free parameter $\lambda=\mp 3\omega^2\ (0\leq \omega <+\infty)$. The obtained geodesics of test particles in the AdS$_4$ spacetime can describe the world lines of the virtual particles and antiparticles in the vacuum. These virtual particle pairs are explained as the result of spontaneous symmetry breaking of the AdS$_4$ spacetime. We also obtain a relation similar to Heisenberg uncertainty relation that describes the quantum vacuum fluctuations.
The geodesics of test particles in the dS$_4$ spacetime can describe the world lines of galaxies moving apart with positive accelerations in the empty universe, and with the red shift.

\section{Acknowledgment}
We thank Phong Pham (HUST) for assistance with preparing the manuscript.



\bibliographystyle{unsrt}
\bibliography{mnsc}

\begin{thebibliography}{10}

\bibitem{padmanabhan2003}
T.~Padmanabhan.
\newblock {\em Phys. Rept.}, 380:235, 2003.

\bibitem{carroll1992}
S.~M. Carroll, W.~H. Press, and E.~L. Turner.
\newblock {\em Annu. Rev. Astron. Astrophys.}, 30:499, 1992.

\bibitem{weinberg1989}
S.~Weinberg.
\newblock {\em Rev. Mod. Phys.}, 61:1, 1989.

\bibitem{carroll2001}
S.~M. Carroll.
\newblock {\em Living Rev. Rel.}, 4:1, 2001.

\bibitem{peebles2003}
P.~J.~E. Peebles.
\newblock {\em Rev. Mod. Phys.}, 75:559, 2003.

\bibitem{hackmann2015}
E.~Hackmann and C.~L{\"a}mmerzahl.
\newblock {\em gr-qc/1505.07973v1}, 2015.

\bibitem{hackmann2008}
E.~Hackmann and C.~L{\"a}mmerzahl.
\newblock {\em Phys. Rev. Lett.}, 100:171101, 2008.

\bibitem{hackmann2010}
E.~Hackmann.
\newblock PhD thesis, Bremen University, 2010.

\bibitem{cruz2005}
N.~Cruz, M.~Olivares, and J.~R. Villanueva.
\newblock {\em Class. Quantum Grav.}, 22:1167, 2005.

\bibitem{bengtsson1998}
I.~Bengtsson.
\newblock {\em www.fysik.su.se/$\sim$ingemar/Kurs.pdf}, 1998.

\bibitem{burdyuzha2013}
V.~Burdyuzha.
\newblock {\em J. of Modern Phys.}, 4:1185, 2013.

\bibitem{ohanian1994}
H.~C. Ohanian and R.~Ruffini.
\newblock {\em Gravitation and spacetime, 2nd ed.}
\newblock W. W. Norton \& Company, Inc., NY, 1994.

\bibitem{ripken2013}
A.~C. Ripken.
\newblock Bachelor thesis, Radboud University Nijmegen, 2013.

\bibitem{moffat2009}
J.~W. Moffat.
\newblock {\em Perimeter Institute}, PIRSA Number: 09050076, 2009.

\bibitem{oldershaw2009}
R.~L. Oldershaw.
\newblock {\em gen-ph/0901.3381}, 2009.

\bibitem{marsh2008}
G.~Marsh.
\newblock {\em gr-qc/0711.0220v2}, 2008.

\end{thebibliography}


\end{document}